\documentclass[fleqn,twoside]{article}
\usepackage{espcrc2}
\usepackage{epsfig}

\def\beq{\begin{equation}}
\def\eeq{\end{equation}}
\def\ts{\left(}
\def\td{\right)}

\usepackage{graphicx}
\usepackage[figuresright]{rotating}

\newcommand{\AmS}{{\protect\the\textfont2
  A\kern-.1667em\lower.5ex\hbox{M}\kern-.125emS}}

\title{
\vspace{-7mm}
\thispagestyle{empty}
\rightline{\small CERN-TH/2001-282}
\vspace{2mm}
Finite-size scaling of interface free energies in the $3d$
  Ising model}

\author{Michele Pepe\address{Laboratoire de Physique Th\'eorique
    Universit\'e de Paris-Sud, B\^atiment 210 F-91405 Orsay-Cedex}
  \thanks{Speaker at the Conference} and 
      Philippe de Forcrand\address{Institut f\"ur Theoretische
    Physik, ETH H\"onggerberg, CH-8093 Z\"urich, 
    Switzerland}$^,$\address{CERN, Theory Division, CH-1211 Gen\`eve 23, 
    Switzerland} }
        
\begin{document}

\begin{abstract}
We perform a study of the universality of the finite size scaling
functions of interface free energies in the $3d$ Ising model. 
Close to the hot/cold phase transition, we observe very good
agreement with the same scaling functions of the $4d$ $SU(2)$
Yang--Mills theory at the deconfinement phase transition.
\end{abstract}

\maketitle

\section{Introduction}
Finite Size Scaling (FSS) is a very powerful tool, particularly useful
in extracting numerical estimates from Monte Carlo
simulations. Numerical simulations give results on systems of
finite size, where phase transitions cannot occur. Thus, a fundamental
task in Monte Carlo studies of phase transitions is the extraction of
results holding in the thermodynamic limit from data obtained in
systems of finite size.
FSS describes how a system of finite size approaches criticality in
the thermodynamic limit.
One of its most far--reaching aspects
concerns the universality of the FSS functions. If universality
holds, complex systems share the same critical behaviour as 
simpler systems. This opens the possibility to use simple models to
obtain accurate information on much more complicated theories.\\
In a recent paper\cite{PhLvS}, a study of the FSS of spatial 't~Hooft loops of
maximal size across the deconfinement phase transition has been
performed in the $SU(2)$ Yang--Mills theory. By this analysis, a
numerical estimate of the dual string tension - for the first
time observed in \cite{tHloop} - has been obtained. Moreover, also the FSS of the
electric free energies of the theory has been investigated.\\
According to the Svetitsky--Yaffe conjecture\cite{SY}, the $4d$
$SU(2)$ Yang--Mills theory at the deconfinement transition is believed
to be in the same universality class as the $3d$ Ising model at the
hot/cold phase transition. 
Several numerical studies have
confirmed the validity of this conjecture. Thus, also
the universal FSS functions of the electric and magnetic free energies of
the $4d$ $SU(2)$ theory measured in \cite{PhLvS}, should match with 
FSS functions observable in the $3d$ Ising model. 
We investigate the universality of the FSS functions in the $3d$ Ising
model at the hot/cold phase transition. In particular we focus on the
FSS of interface free energies.  

\section{The observables}
Our aim is to compare FSS functions measured in the Ising model
with those observed in $SU(2)$ for the magnetic and the electric
free energies.
In order to set this connection, it is useful to recall the
idea behind the Svetitsky--Yaffe conjecture for the case we are
considering now. If we could integrate the degrees of freedom of the
$4d$ $SU(2)$ gauge theory to write an effective action for the
Polyakov loop $P(\vec {x})$, this
effective action would be a $3d$ spin system with symmetry $Z_2$. Then
Svetitsky and Yaffe give arguments according to which the
(complicated) effective action for the Polyakov loop reduces - close
to criticality - to the nearest--neighbour Ising interaction.
If we now consider an $SU(2)$ Yang--Mills theory on the lattice, a
spatial 't~Hooft loop in the $xy$ plane can be obtained by inverting
the sign of the coupling of a set of time--like $zt$ plaquettes. 
In \cite{PhLvS} it is shown that 
$P(\vec {x}) P^{\dagger}(\vec{x}+L\hat{z}) =-1$, where $L$ is the
extension of the system in direction $z$ and $\hat{z}$ is
the unit vector along the $z$ axis. 
This condition can be accomplished in the effective Ising model
for the Polyakov loop, by changing the boundary conditions from
periodic (p.b.c.) to antiperiodic (a.p.b.c.) in direction $z$. 
This generates a frustration in the spin system, which, in the cold
phase, creates a surface of defects. One can set a.p.b.c. in one, two or in all three
directions. In the $SU(2)$ theory, this corresponds to 
switching on 1, 2 or 3 orthogonal spatial 't~Hooft loops of maximal size.

\section{The computation}

We study the $3d$ Ising model on a cubic lattice of size $L^3$,
with the usual ferromagnetic nearest-neighbour interaction 
${\cal{S}}=-\beta \sum_{\vec{x}\mu} \sigma_{\vec{x}}
\sigma_{\vec{x}+\hat{\mu}}$. $\sigma_{\vec{x}}$ is the spin at the
lattice site $\vec{x}$, $\hat{\mu}$ is the unit vector in direction
$\mu$ and $\beta > 0$ is the coupling. The data presented in this
paper have been collected on lattices of size up to $L=32$. More
details about the simulation parameters will be given in a
forthcoming paper. We consider p.b.c. and
a.p.b.c. in $1$, $2$ and $3$ directions.
${\cal{Z}}_k (i)$, with respectively $i=0,1,2,3$ are the partition
functions for these 4 different choices. The observable we consider
is the free energy cost $F_k (i)$ to create such interfaces. It
is given by
\beq\label{magsec}
F_k (i) \equiv -\log \frac{{\cal{Z}}_k (i)}{{\cal{Z}}_k (0)}
\;\;\;\; i=1,2,3
\eeq
The counterparts of the electric free energies of the $SU(2)$ theory are obtained
by performing the $Z_2$ Fourier transform \cite{tHooft} in (\ref{magsec}).
In terms of the ${\cal{Z}}_k (i)$, their expressions are
\begin{eqnarray}\label{elsec}
Z_e (1) \hskip-3mm&=& \hskip-3mm
\frac{1}{{\cal{N}}_e} \hskip-1mm \ts {\cal{Z}}_k (0) + {\cal{Z}}_k (1) 
- {\cal{Z}}_k (2) - {\cal{Z}}_k (3)\td  \;\;\\
Z_e (2) \hskip-3mm&=& \hskip-3mm
\frac{1}{{\cal{N}}_e} \hskip-1mm \ts {\cal{Z}}_k (0) - {\cal{Z}}_k (1)
- {\cal{Z}}_k (2) + {\cal{Z}}_k (3)\td  \;\;\\
Z_e (3) \hskip-3mm&=& \hskip-3mm
\frac{1}{{\cal{N}}_e} \hskip-1mm \ts {\cal{Z}}_k (0) - 3 {\cal{Z}}_k (1)
+ 3 {\cal{Z}}_k (2) - {\cal{Z}}_k (3)\td\;\;
\end{eqnarray}
where ${\cal{N}}_e={\cal{Z}}_k (0) +3 {\cal{Z}}_k (1)
+3 {\cal{Z}}_k (2) + {\cal{Z}}_k (3)$. 
The correlation length $\xi$ sets the scale of the distances.
It has a critical behaviour with exponent $\nu\approx 0.63$: 
$\xi\sim |t|^{-\nu}$, where $t=1-\beta_c/\beta$. 
We express the FSS functions in terms of the scaling variable 
$x=\mbox{sign}(t) L |t|^{\nu} \propto L/\xi$. 
In Figure~1, we compare the FSS functions of $F_k (1)$ between the
Ising model and $SU(2)$. We show results both in the
cold/deconfined phase ($x>0$) and the hot/confined one ($x<0$). 
On one hand, we observe that,
according to FSS expectations, the data collected at various values of
$L$ and $\beta$ lie on a single curve, depending only on 
$x$. On the other hand, it turns out to agree excellently
with the FSS function of one spatial 't~Hooft loop in $SU(2)$. Note
that the scaling functions for $SU(2)$ and the 
Ising model do not directly match from the raw data. 
A rescaling of the correlation length by a factor $\alpha$ is
necessary: $\xi_{\mbox{\tiny{Ising}}} = \xi_{\tiny{SU(2)}}/\alpha$.
This implies that the variable $x$ defined for the Ising model and
the one used in $SU(2)$ \cite{PhLvS} are related by: 
$x_{\mbox{\tiny{Ising}}} = \alpha x_{\tiny{SU(2)}}$. We have estimated the
value $\alpha=1.88(2)$ by rescaling the $x>0$ Ising data in such a way that
they overlap with the $SU(2)$ ones. We stress that, once $\alpha$
has been estimated, its value remains fixed and it is no more a
fitting parameter. In Figures 2 and 3
the $x$ variable for the $SU(2)$ data has been rescaled by $\alpha$. 
\begin{figure}[tb] 
\hspace*{-.4cm}
\epsfig{file=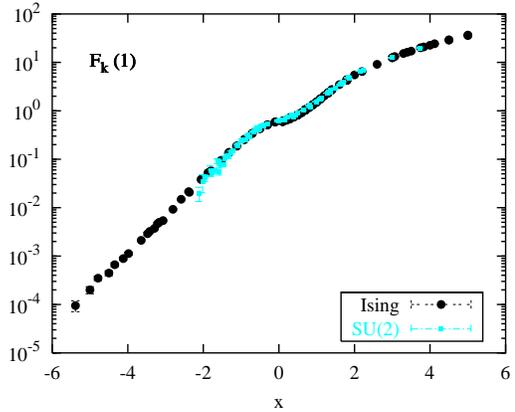,width=80mm,angle=0,clip}
\vspace*{-1.5cm}
\caption{Comparison of the FSS functions of $F_k (1)$ between the Ising
  model and $SU(2)$.}
\vskip -1.cm
\end{figure} 
In Figure 2, we display the comparison Ising--$SU(2)$ for $F_k (i)$ 
$i=1,2,3$ at $x>0$. 
The interface tension $\sigma_i$ is a fitting parameter only for 
$F_k (1)$: we estimate the value $\sigma_1 =1.495(15)$. 
For a full comparison, we have used
the same fitting ansatz as in \cite{PhLvS}.
This result is consistent with -- but more precise than --  the value
measured in \cite{HP}. For $F_k (i)$ $i=2,3$, we have set 
$\sigma_i =  \sigma_1\sqrt{i}$, as follows from the expectations of
the minimal interface area. This hypothesis is well
satisfied, as apparent from the good quality of the fits. The very
good matching Ising--$SU(2)$ of the FSS functions for the free energy
interface holds also for the cases of a.p.b.c. in 2 and 3 directions. 
\begin{figure}[tb] 
\hspace*{-.4cm}
\epsfig{file=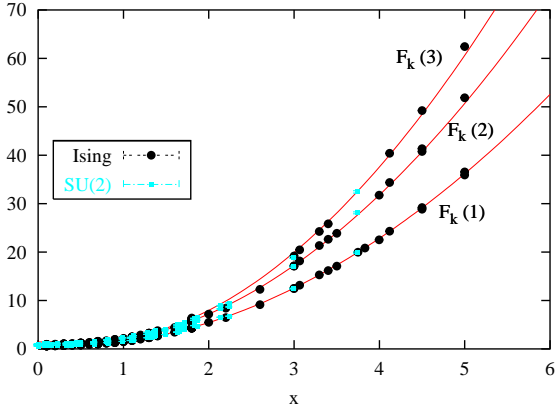,width=80mm,angle=0,clip}
\vspace*{-1.5cm}
\caption{Comparison Ising-$SU(2)$ of the FSS functions of $F_k (i)$,
$i=1,2,3$.
The continuous lines are
fits according to the ansatz used in \protect\cite{PhLvS}.}
\vskip -.95cm
\end{figure} 
Figure~3 concerns the FSS for $F_e (1)$; similar results hold for 
$F_e (2)$ and $F_e (3)$. Thus, also for the electric free energies, we
observe very good agreement between the FSS functions of the
two theories, with much increased accuracy in the Ising case. 
Finally, we have measured the product $\sigma_1 \xi_+^2$, where
$\xi_+$ is the correlation length in the hot phase. At
large $x$ and close to the critical point, the FSS functions of $F_k (1)$ and 
$F_e (1)$ have the simple asymptotic forms:
$F_k (1) \approx \sigma_1 x^2 +C_k$ and 
$F_e (1) \approx |x|/\xi_{+} +C_e$, where $C_k$ and $C_e$ are two constants. 
Then at large $|x|$, we have 
$F_k(1)/F_e(1)^2 \rightarrow \sigma_1 \xi_+^2$.
In order for this ratio to reach its asymptotic behaviour
sooner, we have subtracted a constant $C$ to the interface free
energy $F_k (1)$. 
In Figure~4 we plot 
$A\equiv (F_k(1)-C)/F_e(1)^2$ as function of $|x|$.
The data clearly show a flattening to a constant at large $|x|$.
Fitting the last 5 points to the right with a constant, we obtain 
$(\sigma_1 \xi_+^2) = 0.44(1)$.
Extended discussions and details about the shown results 
will be presented in a forthcoming paper.
\begin{figure}[tb] 
\hspace*{-.4cm}
\epsfig{file=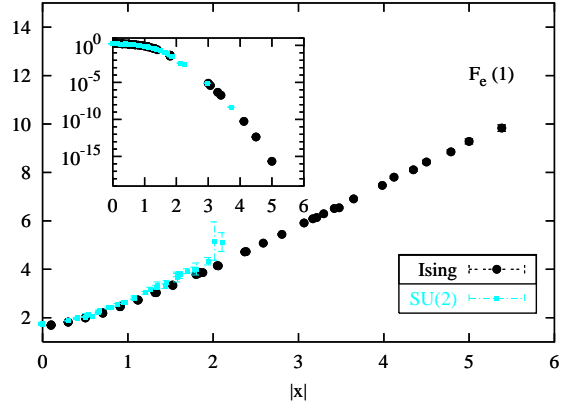,width=80mm,angle=0,clip}
\vspace*{-1.5cm}
\caption{Comparison Ising-$SU(2)$ of the FSS functions of
$F_e (1)$. The data in the main figure refer to the hot/confined
phase. The insert displays the results in the cold/deconfined phase.} 
\vskip -.8cm
\end{figure} 
\begin{figure}[tb] 
\hspace*{-.4cm}
\epsfig{file=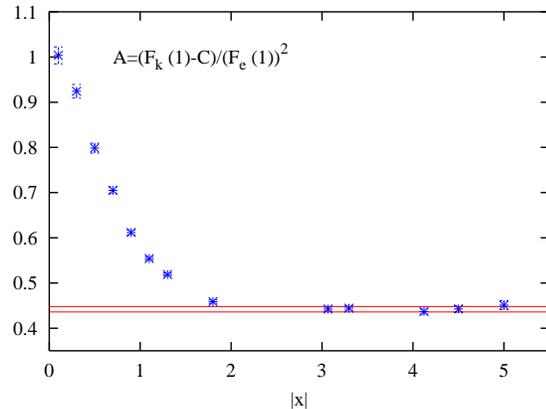,width=80mm,angle=0,clip}
\vspace*{-1.5cm}
\caption{Approach of $A$ to a constant.
The strip is the estimated asymptotic value 
$\sigma_1 \xi_+^2$.} 
\vskip -.9cm
\end{figure} 

\noindent
{\em{Acknowledgments}}. We gratefully acknowledge useful discussions
with M.~Caselle, F.~Gliozzi, M.~Hasenbusch, G.~M\"unster, C.~Roiesnel,
T.~Tomboulis and L.~von~Smekal. We also thank V.~Antonelli and M.~Comi for technical
support. M.P. is supported by the European Community's Human potential 
programme under HPRN-CT-2000-00145 Hadrons/LatticeQCD.


\begin{thebibliography}{9}
\bibitem{PhLvS}
P.~de Forcrand and L.~von Smekal, hep-lat/0107018 and hep-lat/0110135.
\bibitem{tHloop}
P.~de Forcrand, M.~D'Elia and M.~Pepe, Phys.\ Rev.\ Lett.\  {\bf 86}
(2001) 1438. 
\bibitem{SY}
B.~Svetitsky and L.~G.~Yaffe, Nucl.\ Phys.\ B {\bf 210} (1982) 423.
\bibitem{tHooft}
G.~'t Hooft, Nucl.\ Phys.\ B {\bf 138} (1978) 1.
\bibitem{HP}
M.~Hasenbusch and K.~Pinn, Physica A {\bf 245} (1997) 366.
\end{thebibliography}
\end{document}